\documentstyle[aps,multicol,epsf]{revtex}
\begin{document}
\draft
\title{Depinning of an anisotropic interface in random media: The tilt effect \\}
\author{K.-I.~Goh$^{1}$, H. Jeong$^{2}$, B. Kahng$^{3}$ and D.~Kim$^{1}$}
\address{
1. Center for Theoretical Physics and Department of Physics, 
Seoul National University, Seoul 151-742, Korea \\
2. Department of Physics, University of Notre Dame, Notre Dame, IN 46556 \\ 
3. Department of Physics and Center for Advanced Materials and Devices,
Konkuk University, Seoul 143-701, Korea \\}
\date{\today}
\maketitle
\thispagestyle{empty}

\begin{abstract}
We study the tilt dependence of the pinning-depinning 
transition for an interface 
described by the anisotropic quenched Kardar-Parisi-Zhang equation 
in 2+1 dimensions, where the two signs of the nonlinear terms are 
different from each other. 
When the substrate is tilted by $m$ along 
the positive sign direction, the critical force $F_c(m)$ depends on 
$m$ as $F_c(m)-F_c(0) \sim -|m|^{1.9(1)}$.
The interface velocity $v$ near the critical force follows the scaling form 
$v \sim |f|^{\theta}\Psi_{\pm}(m^2 /|f|^{\theta+\phi})$ 
with $\theta = 0.9(1)$ and $\phi= 0.2(1)$, where $f \equiv F-F_c(0)$ and
$F$ is the driving force.
\end{abstract}

\pacs{PACS numbers: 68.35.Fx, 05.40.+j, 64.60.Ht}

\begin{multicols}{2}
\narrowtext

The pinning-depinning (PD) transition
from a pinned to a moving state 
is of interest due to its relevance to many physical systems.
Typical examples include interface growth in porous (disordered)
media under external pressure~\cite{porous,kardar},
dynamics of a domain wall under external field
~\cite{rfield,bausch,narayan}, 
dynamics of a charge density wave under external field
~\cite{cdw,fukuyama}, 
and vortex motion in superconductors under external 
current~\cite{ertas,hwa}. In the PD transition, there exists 
a critical value, $F_c$, of the external driving force $F$, 
such that for $F<F_c$ the interface (or charge, or vortex) is pinned 
by the disorder, while for $F>F_c$, it moves forward with a constant
velocity $v$, leading to a transition across $F_c$.
The velocity $v$ plays the role of the order parameter and typically behaves as  
\begin{equation}\label{vf}
v \sim (F-F_c)^{\theta} 
\end{equation}
with the velocity exponent $\theta$. \\ 

The interface dynamics in disordered media may be
described via the Langevin-type continuum equation for the interface position $h(x,t);$ 
\begin{equation}\label{qew}
\partial_t h(x,t) = {\cal{K}}[h] + F + \eta(x,h).
\end{equation}
The first term on the right hand side of Eq.~(\ref{qew}) 
describes the configuration dependent force, the second is the external driving force, 
and the last, the quenched random noise, independent of time, describes the 
fluctuating force due to randomness or impurities in the medium. 
The random noise is assumed to have the properties, 
$\langle \eta(x,h) \rangle =0$
and
$\langle \eta(x,h)\eta(x',h')\rangle =2D \delta^{d}(x-x')\delta(h-h'),$
where the angular brackets represent the average over different 
realizations and $d$ is the substrate dimension.
When ${\cal K}[h]=\nu \nabla^2 h,$ the resulting linear equation is called the quenched 
Edwards-Wilkinson equation (QEW)~\cite{ew}.\\

Recently, a couple of stochastic models mimicking the interface dynamics 
in disordered media have been introduced \cite{boston,tang}, 
displaying the PD transition different from the QEW universality 
behavior~\cite{abs}. It was proposed that the models are described 
by the Kardar-Parisi-Zhang equation with quenched noise (QKPZ)~\cite{kpz}, where
\begin{equation}\label{qkpz}
{\cal K}[h]=\nu{\nabla^2 h }+{\lambda \over 2}( \nabla h)^2. 
\end{equation} 
The nonlinear term comes from the anisotropic nature of disordered 
medium, thus non-vanishing at the critical force $F_c$ as shown by 
Tang {\it et al.} \cite{dhar} in the context of vortex dynamics. 
When effective pinning force in the random impurity takes the form
$(\Delta_h+s^2\Delta_x)^{2/3}/(1+s^2)^{2/3}$, where $s \equiv \partial_x h$ 
is the local slope and $\Delta_h^{1/2}$ and $\Delta_x^{1/2}$ are
the amplitudes of random forces in the $h$ and $x$ directions, 
respectively, the nonlinear term is derived by 
expanding the random force in power of the slope $s$, 
leading to $\lambda \propto (\Delta_h - \Delta_x)$. 
Therefore, when $\Delta_h > \Delta_x$ ($\Delta_h < \Delta_x$) 
i.e., when the interface is driven along the hard (easy) direction, 
$\lambda$ is positive (negative), and when the medium is isotropic, 
$\Delta_h =\Delta_x$, the nonlinear term vanishes.  
It has been shown \cite{jeong,jeong2} that the interface dynamics 
of the QKPZ equation depends on the sign of $\lambda$, 
in contrast to the thermal case where the sign is irrelevant \cite{kpz}.\\

The critical behavior of the PD transition for the QKPZ 
equation has been thoroughly studied in 1+1 dimensions. 
For $\lambda > 0$, the PD transition is continuous, 
and the interface at $F_c$ is characterized in terms of 
the directed percolation (DP) cluster \cite{dp} spanning 
in the perpendicular-to-the-growth direction\cite{boston,tang}.
In this case, the effective nonlinear coefficient
diverges as $\sim (F-F_c)^{-\phi}$
as $F \rightarrow F_c^+$,
and the critical force $F_c$ depends 
on the substrate-tilt $m$ as 
\begin{equation}\label{fcm}
F_c(m)-F_c(0) \sim -|m|^{1/\nu(1-\alpha)}, 
\end{equation}
where $\nu$ and $\alpha$ are the correlation length and the roughness 
exponent, respectively. 
These exponents are related to one another as~\cite{dhar} 
\begin{equation}\label{relation}
\phi=2\nu(1-\alpha)-\theta. 
\end{equation}
For $\lambda <0$, the surface at $F_c$ forms a facet with a 
characteristic slope $s_c$. The effective nonlinear 
coefficient is insensitive to $F$. 
When the substrate-tilt $m$ is smaller than $s_c$, 
the PD transition is discontinuous, 
and $F_c$ is independent of $m$, while the transition is 
continuous and $F_c$ increases with $m$ for $m > s_c$. 
The discontinuous transition is caused by the presence of 
a critical pinning force due to both the negative nonlinear term 
and random noise, which is localized. Once this pinning force 
is overcome by increasing external force $F$, the surface moves
forward abruptly, yielding the velocity jump at $F_c$. 
The amount of the velocity jump decreases
with increasing $m$, and vanishes at the characteristic
tilt $m_c=s_c$. For $m > m_c$, the velocity increases
continuously from zero, and the PD transition is continuous.
Accordingly, the characteristic substrate-tilt $m_c$ is a 
multicritical point \cite{jeong,jeong2}.\\

Since the sign of the nonlinear term is relevant in the quenched
case, it would be interesting to consider the case of the 
anisotropic QKPZ equation in 2+1 dimensions, where the 
signs of the nonlinear terms are alternative. Thus, we consider  
\begin{eqnarray}\label{aqkpz}
\partial_t h=\nu_x {\partial}_x^2 h
+ \nu_y {\partial}_y^2 h +{\lambda_x \over 2}(\partial_x h)^2
+{\lambda_y \over 2}(\partial_y h)^2 \nonumber \\
+F+\eta(x,y,h),
\end{eqnarray}
where $\lambda_x >0$ and $\lambda_y <0$.
The anisotropic case is in particular 
interesting due to its application to the vortex motion in 
disordered system \cite{hwa} and the adatom motion on step edge 
in epitaxial surface \cite{wolf}.  
It has been shown \cite{wolf} that for the thermal case, 
the two nonlinear terms cancel each other effectively, 
thereby the anisotropic KPZ equation is reduced to the linear equation, 
the EW equation. For the quenched case, however,  
the interface dynamics in each direction are different from 
each other and the surface morphology is anisotropic: the surface is  
gently sloping in the positive sign direction 
($x$-direction), and is of the shape of a mountain range with steep slope 
in the negative sign direction ($y$-direction). 
In spite of the facet shape in the negative sign direction, the PD 
transition is continuous due to the critical behavior in 
the positive sign direction. Consequently, one may expect 
that the critical force and the effective nonlinear coefficient 
along the positive sign direction can be described by the scaling 
theory introduced for $\lambda > 0$ in 1+1 dimensions.  
In this Brief Report, we show, from extensive numerical simulations, 
that this is indeed the case and determine the scaling exponents
$\phi$ and $\nu(1-\alpha)$ independently. \\

\begin{figure}
\centerline{\epsfxsize=9.0cm \epsfbox{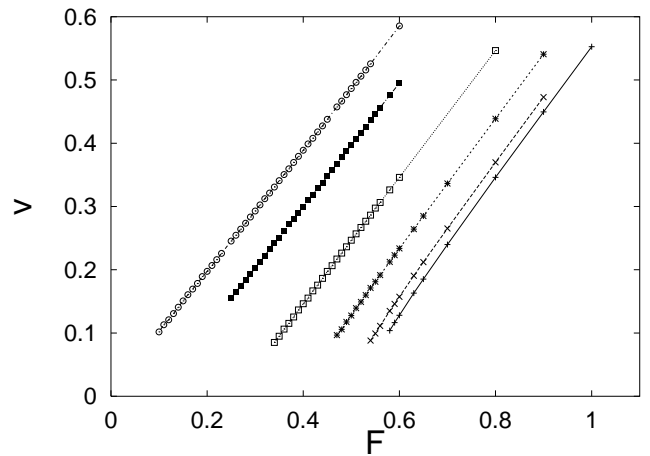}}
\caption{The interface velocity as a function of $F$ for various $m$.
The curves correspond to the cases of $m =$ 0.0, 0.2, 0.4, 0.6, 0.8,
0.9, from right to left.}
\label{vmf}
\end{figure}

Direct numerical integration has been carried out using   
standard discretization techniques \cite{vicsek,leschhorn}, 
in which we choose the parameters, $\nu_x=\nu_y=1$, 
$\lambda_x=-\lambda_y=1$, and a temporal increment 
$\Delta t=0.01$. The noise is discretized 
as $\eta (x,y,[h])$, where $[\cdots]$ means 
the integer part, and $\eta$ is uniformly distributed 
in $[-a/2,a/2]$ with $a=(10)^{2/3}$.
In order to consider the tilt-dependence, we tilted the
substrate as $h(x,y,0) = mx$ along the positive sign direction,
and used the helicoidal boundary condition, $h(L+x,y,t) = h(x,y,t)+Lm$.
We measured the growth velocity as a function of the external force $F$
for several values of substrate-tilt $m$, which is shown in 
Fig.~\ref{vmf}. For $m=0$, we found that $\theta = 0.9(1)$ as in 
Ref.~\cite{jeong}. But for non-zero $m$, the exponent $\theta(m)$
is generically 1, different from its $m=0$ value~\cite{dhar}.
This picture can be seen in Fig.~\ref{vmf}, which shows straight
lines for large $m$. \\ 

\begin{figure}
\centerline{\epsfxsize=9.0cm \epsfbox{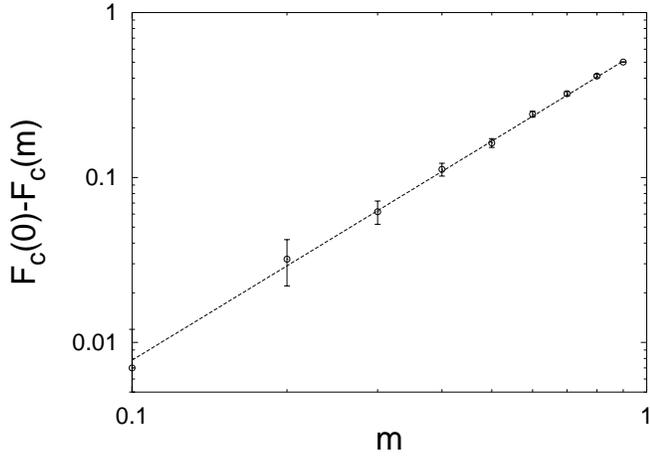}}
\caption{ Log-log plot of $F_c(0)-F_c(m)$ versus $m$.
The dashed line has slope 1.9, drawn for the eye.
}
\label{fcmpa}
\end{figure}

The critical force $F_c$ is estimated as the maximum value
of $F$ for which all samples of 10 are pinned until 
a large Monte Carlo steps, typically  $10^5 \Delta t$. 
In this way, we find $F_c(0) \approx 0.51(1)$.
This value is slightly larger than that obtained by
extrapolating the velocity curve, $\approx 0.50$. 
Also, we estimated the critical force $F_c(m)$ as a function
of the substrate-tilt $m$. The critical force $F_c(m)$
decreases with increasing substrate-tilt $m$ with the exponent
$1/\nu(1-\alpha)\approx 1.9(1)$ as shown in Fig.~\ref{fcmpa}.\\

\begin{figure}
\centerline{\epsfxsize=9.0cm \epsfbox{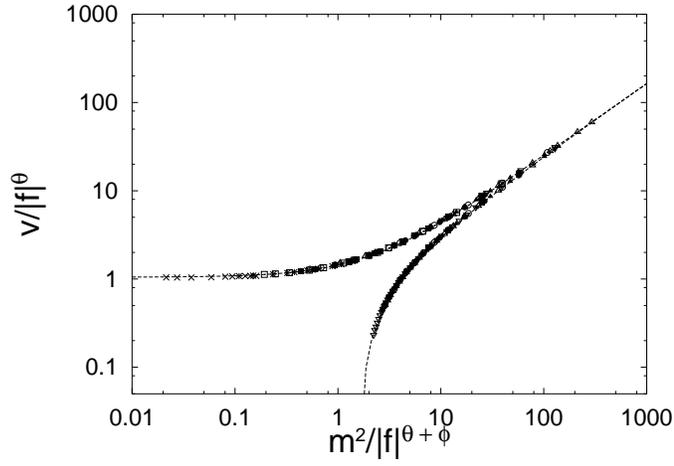}}
\caption{Data collapse for the interface velocity using Eq.~(\ref{scaling})
with the exponents
$\theta=0.9(1)$ and $\phi=0.2(1)$. The dashed lines are drawn for the 
approximate form of the scaling functions, Eqs.~(\ref{psi+}) and (\ref{psi-}).}
\label{collapse}
\end{figure}

The PD transition is continuous due to the positive nonlinear
term. 
To determine the exponent $\phi$ independently, we assume
the scaling form for the interface velocity $v(F,m)$
as~\cite{abs,dhar}
\begin{equation}\label{scaling}
v \sim |f|^{\theta} \Psi_{\pm}\left( {m^2 \over |f|^{\theta+\phi}}\right).
\end{equation}
Here, $f\equiv F-F_c(0)$ and the subscript $+(-)$ denotes the positive (negative)
$f$ branch. 
To be consistent with Eqs.~(\ref{vf}) and (\ref{fcm}) and the
fact that $\theta(m)=1$ for $m\neq0$,
the scaling function should behave as,
\begin{eqnarray}
&&\Psi_{\pm}(x\rightarrow\infty) \sim x^{\theta/(\theta+\phi)}, \nonumber \\
&&\Psi_+(0^+) = {\mathrm constant}, \nonumber
\end{eqnarray}
and for some positive constant $x_0$, 
\begin{eqnarray}
&&\Psi_-(x_0)= 0,   \nonumber\\
&&\Psi'_-(x_0)={\mathrm constant},\nonumber 
\end{eqnarray}
where the prime denotes the derivative with respect to $x$.
We find that the velocity data $v(F,m)$ near the transition 
can be collapsed onto a single curve consistent with the scaling form. 
The best collapse is achieved with
$F_c(0) \approx 0.50$ and the exponents $\theta \approx 0.9(1)$ 
and $\phi \approx 0.2(1)$, as shown in Fig.~\ref{collapse}. 
From a simple fit to the data, we obtain an approximate functional form 
of the scaling function $\Psi_{\pm}(x)$ as 
\begin{equation}\label{psi+}
\Psi_{+}(x) \approx A(x+B)^{{\theta}/{(\theta+\phi)}}
\end{equation}
and,
\begin{equation}\label{psi-}
\Psi_{-}(x) \approx A(x^{{\theta}/{(\theta+\phi)}}-C),
\end{equation}
with constants $A\approx0.57(1)$, $B\approx 2.1(1)$,  
and $C=x_0^{\theta/(\theta+\phi)}\approx 1.5(1)$. 
Here we assume that $\Psi_+(x)$ is analytic at $x=0.$
In Fig.~\ref{collapse},
we also plot Eqs.~(\ref{psi+}) and (\ref{psi-}) with dashed lines.
The exponents thus obtained satisfy the scaling relation, Eq.~(\ref{relation}), 
within the errors.\\ 

\begin{figure}
\centerline{\epsfxsize=9.0cm \epsfbox{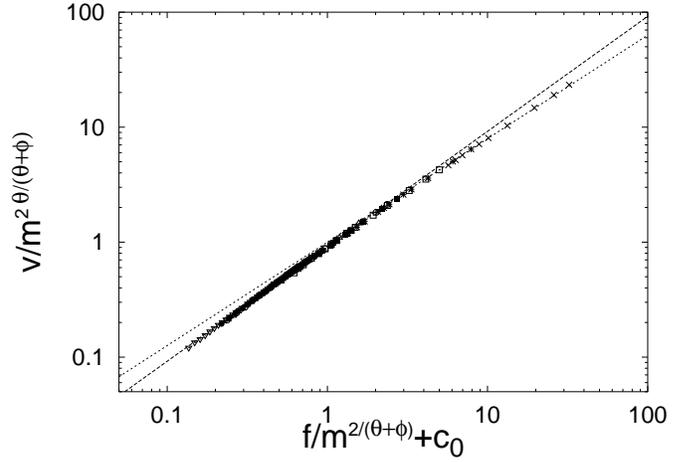}}
\caption{Data collapse for the interface velocity using Eq.~(\ref{altereq})
with $\theta=0.9(1)$ and $\phi=0.2(1)$. The dotted (dashed) line has slope 
0.9 (1.0), drawn for the eye.} 
\label{alterfig}
\end{figure}

Alternatively, one may put the scaling form as 
\begin{equation}\label{altereq}
v \sim m^{2\theta/(\theta+\phi)}\Phi\left({f\over{m^{2/(\theta+\phi)}}}\right),
\end{equation}
with
\begin{eqnarray}
&&\Phi(x\rightarrow\infty) \sim x^{\theta},\nonumber \\
&&\Phi(-c_0)=0,\nonumber \\
&&\Phi'(-c_0)= {\mathrm constant},\nonumber
\end{eqnarray}
where the positive constant $c_0$ is related to $x_0$ 
via $c_0 = x_0^{-1/(\theta+\phi)}$. With this form, the data can be described 
by a single scaling function $\Phi(x)$. 
The scaling plot using the scaling form 
Eq.~(\ref{altereq}) is shown in Fig.~\ref{alterfig}. There, we have shifted 
the argument $x$ of the scaling function $\Phi(x)$ by $c_0$ 
to make the argument positive. The slope of 
the log-log plot of the scaling function shows a crossover from 0.9 to 1.0
with increasing the substrate-tilt $m$, which confirms the fact that 
$\theta(m) = 1$ for $m\neq0$. 
\\

In summary, we have investigated the critical behaviors of the 
tilted anisotropic QKPZ equation at the PD transition. 
The PD transition is continuous when it is tilted
along the positive sign direction. 
It is shown that the data can be collapsed onto a single
curve, with the exponents $\theta=0.9(1)$ and $\phi=0.2(1)$.
The functional form of the scaling function is also numerically
determined.  
We have measured, independently, the exponent which describes 
the variation of $F_c$ with respect to $m$, and confirmed their 
consistency.  \\ 

This work was supported in part by the Korean Research Foundation 
(1999-015-D10070).\\

\end{multicols}
\end{document}